\documentclass[vecphys]{svmult}

\usepackage{makeidx}         
\usepackage{graphicx}        
\usepackage{multicol}        
\usepackage[bottom]{footmisc}
\usepackage{color}

\makeindex             

\newcommand       \beq          {\begin{equation}}
\newcommand       \eeq          {\end{equation}}

\newcommand       \Angstrom     {\,{\rm \AA}}

\newcommand       \cm           {\,{\rm cm}}

\newcommand       \s            {\,{\rm s}}

\newcommand       \simlt        {<}

\newcommand       \simali       {\sim\,}

\newcommand       \mum          {\,{\rm \mu m}}

\newcommand       \Io           {I_{\rm o}}
\newcommand       \Cgeo         {C_{\rm geo}}
\newcommand       \Csca         {C_{\rm sca}}
\newcommand       \Cabs         {C_{\rm abs}}
\newcommand       \Cext         {C_{\rm ext}}
\newcommand       \Cpr          {C_{\rm pr}}
\newcommand       \Qsca         {Q_{\rm sca}}
\newcommand       \Qabs         {Q_{\rm abs}}
\newcommand       \Qext         {Q_{\rm ext}}
\newcommand       \Qpr         {Q_{\rm pr}}
\newcommand	\bx	{\vec{\bf x}}
\newcommand	\bE	{\vec{\bf E}}
\newcommand	\bH	{\vec{\bf H}}
\newcommand	\bB	{\vec{\bf B}}
\newcommand	\bD	{\vec{\bf D}}
\newcommand	\bJ	{\vec{\bf J}}

\newcommand	\dotx	{\dot{\vec{\bf x}}}
\newcommand	\ddotx	{\ddot{\vec{\bf x}}}
\newcommand	\dipmom	{\vec{\bf p}} 
\newcommand	\pol	{\vec{\bf P}}
\newcommand	\balpha	{\vec{\bf \alpha}}

\newcommand     \omegao {{\omega_{\rm o}}}
\newcommand     \omegap {{\omega_{\rm p}}}
\newcommand     \omegape {{\omega_{{\rm p},e}}}
\newcommand     \epsilonre {{\varepsilon^{\prime}}}
\newcommand     \epsilonim {{\varepsilon^{\prime\prime}}}
\newcommand     \mre       {{m^{\prime}}}
\newcommand     \mim       {{m^{\prime\prime}}}

\newcommand     \cext      {C_{\rm ext}}
\newcommand     \cabs      {C_{\rm abs}}
\newcommand     \csca      {C_{\rm sca}}

\newcommand       \keV           {\,{\rm keV}}

\newcommand       \rhoR         {\rho_{\rm R}}

\newcommand	  \copol	{C_{\rm pol}}
\newcommand	  \cppol	{C_{\rm pol}}
\newcommand	  \cpar		{C_{\rm abs}^{\|}}
\newcommand	  \cper		{C_{\rm abs}^{\bot}} 
\newcommand	  \ahat		{r_{a}}
\newcommand	  \bhat		{r_{b}}
\newcommand	  \chat		{r_{c}}
\newcommand	  \ehat		{\xi_{e}}
\begin{document}

\title*{Optical Properties of Dust}
\author{Aigen Li}
\institute{Department of Physics and Astronomy,
           University of Missouri,\\
           Columbia, MO 65211;
\texttt{LiA@missouri.edu}}
\maketitle

\vspace{-0.2cm}
\begin{abstract}
\vspace{-0.6cm}
Except in a few cases cosmic dust can be studied {\it in situ}
or in terrestrial laboratories,
essentially all of our information concerning 
the nature of cosmic dust depends upon its interaction 
with electromagnetic radiation.
This chapter presents the theoretical basis 
for describing the optical properties of dust 
-- how it absorbs and scatters starlight
and reradiates the absorbed energy at longer wavelengths.
\end{abstract}

\vspace{-8mm}
\section{Introduction\label{sec:intro}}
\vspace{-2mm}
Dust is everywhere in the Universe: 
it is a ubiquitous feature of the cosmos, 
impinging directly or indirectly on 
most fields of modern astronomy. 
It occurs in a wide variety of astrophysical regions,
ranging from the local environment of the Earth to
distant galaxies and quasars: 
from meteorites originated in the asteroid belt, 
the most pristine solar system objects -- comets, 
and stratospherically collected 
interplanetary dust particles 
(IDPs) of either cometary or asteroidal origin
to external galaxies 
(both normal and active, nearby and distant)
and circumnuclear tori around active galactic nuclei; 
from circumstellar envelopes around evolved stars
(cool red giants, AGB stars) and Wolf-Rayet stars, 
planetary nebulae, nova and supernova ejecta, 
and supernova remnants 
to interstellar clouds and star-forming regions; 
from the terrestrial zodiacal cloud 
to protoplanetary disks around young stellar objects 
and debris disks around main-sequence stars ...

Dust plays an increasingly important role in astrophysics.
It has a dramatic effect on the Universe 
by affecting the physical conditions 
and processes taking place within the Universe
and shaping the appearance of dusty objects
(e.g. cometary comae, reflection nebulae, dust disks,
and galaxies) 
(i) as an absorber, scatterer, polarizer,
    and emitter of electromagnetic radiation;
(ii) as a revealer of heavily obscured objects 
     (e.g. IR sources) of which 
     we might otherwise be unaware;
(iii) as a driver for the mass loss of evolved stars;
(iv) as a sink of heavy elements 
     which if otherwise in the gas phase,
     would profoundly affect the interstellar gas chemistry;  
(v) as an efficient catalyst for the formation of H$_2$ 
    and other simple molecules 
    as well as complex organic molecules
    (and as a protector by shielding them 
    from photodissociating ultraviolet [UV] photons) 
    in the interstellar medium (ISM); 
(vi) as an efficient agent for heating the interstellar gas 
     by providing photoelectrons;
(vii) as an important coolant in dense regions 
      by radiating infrared (IR) photons
      (which is particularly important for the process
       of star formation in dense clouds by 
       removing the gravitational energy of
       collapsing clouds and 
       allowing star formation to take place); 
(viii) as an active participant in interstellar gas dynamics
       by communicating radiation pressure from starlight 
       to the gas, and providing coupling of the magnetic field 
       to the gas in regions of low fractional ionization;
(ix) as a building block in the formation 
     of stars and planetary bodies and finally, 
(x) as a diagnosis of the physical conditions 
    (e.g. gas density, temperature, radiation
     intensity, electron density, magnetic field)
     of the regions where dust is seen.

The dust in the space between stars -- interstellar dust --
is the most extensively studied cosmic dust type, 
with circumstellar dust (``stardust''), 
cometary dust and IDPs coming second.
Interstellar dust is an important constituent of
the Milky Way and external galaxies.
The presence of dust in galaxies limits our ability 
to interpret the local and distant Universe because 
dust extinction dims and reddens the galaxy light in
the UV-optical-near-IR windows, where the vast majority
of the astronomical data have been obtained. 
In order to infer the stellar content of a galaxy,
or the history of star formation in the Universe,
it is essential to correct for the effects of
interstellar extinction. 
Dust absorbs starlight and reradiates at longer 
wavelengths. Nearly half of the bolometric luminosity 
of the local Universe is reprocessed by dust into 
the mid- and far-IR. 

Stardust, cometary dust and IDPs are directly or indirectly
related to interstellar dust: stardust, condensed in 
the cool atmospheres of evolved stars or supernova ejecta
and subsequently injected into the ISM, is considered as 
a major source of interstellar dust, 
although the bulk of interstellar dust
is not really stardust but must have recondensed 
in the ISM \cite{Draine_1990}.  
Comets, because of their cold formation and cold storage,
are considered as the most primitive objects 
in the solar system
and best preserve the composition of 
the presolar molecular cloud among all solar system bodies.
Greenberg \cite{Greenberg_1982} argued that  
comets are made of unaltered pristine interstellar materials 
with only the most volatile components partially evaporated,
although it has also been proposed that 
cometary materials have been subjected to evaporation, 
recondensation and other reprocessing in the protosolar 
nebula and therefore have lost all the records of 
the presolar molecular cloud out of which they have formed.
Genuine presolar grains have been identified in
IDPs and primitive meteorites 
based on their isotopic anomalies \cite{Clayton_Nittler_2004},
indicating that stardust can survive journeys
from its birth in stellar outflows and supernova explosions,
through the ISM,
the formation of the solar system,
and its ultimate incorporation into asteroids and comets. 

Except in a few cases cosmic dust can be studied {\it in situ}
(e.g. cometary dust \cite{Kissel_2007}, dust in the local 
interstellar cloud entering our solar system \cite{Grun_2005})
or in terrestrial laboratories 
(e.g. IDPs and meteorites \cite{Clayton_Nittler_2004},
cometary dust \cite{Brownlee_2006}),
our knowledge about cosmic dust  
is mainly derived from its interaction with 
electromagnetic radiation: 
extinction (scattering, absorption), 
polarization, and emission.
Dust reveals its presence and physical and
chemical properties and provides clues 
about the environment where it is found
by scattering and absorbing starlight 
(or photons from other objects) 
and reradiating the absorbed energy
at longer wavelengths.

This chapter deals with the optical properties
of dust, i.e., how light is absorbed, scattered,
and reradiated by cosmic dust.
The subject of light scattering by small particles
is a vast, fast-developing field. 
In this chapter I restrict myself to
astrophysically-relevant topics.
In \S\ref{sec:Phys_Scat} I present a brief summary of
the underlying physics of light scattering.
The basic scattering terms are defined in \S\ref{sec:Scat_Def}.
In \S\ref{sec:Diel_Func} I discuss the physical basis 
of the dielectric functions of dust materials. 
In \S\ref{sec:Scat_Approx} and \S\ref{sec:Scat_Tech}
I respectively summarize the analytic and 
numerical solutions for calculating the absorption 
and scattering parameters of dust.

\begin{figure}
\centering
\vspace{-2mm}
\includegraphics[height=9.5cm]{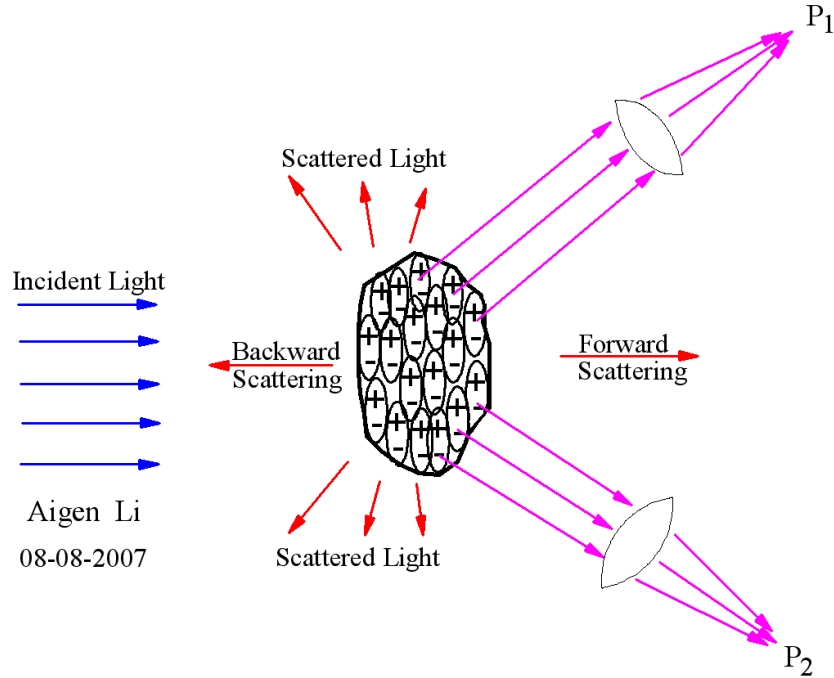}
\vspace{-3mm}
\caption{
         \label{fig:scat1}       
         A conceptual illustration of the scattering 
         of light by dust. The dust is conceptually 
         subdivided into many small regions.
         Upon illuminated by an incident electromagnetic wave,
         in each small region a dipole moment 
         (of which the amplitude and phase 
         depend on the composition the dust)
         will be induced.
         These dipoles oscillate at the frequency of 
         the incident wave and scatter secondary radiation 
         in all directions.
         The total scattered field of a given direction 
         (e.g. $P_1$, $P_2$) is the sum of the scattered 
         wavelets, where due account is taken of their 
         phase differences.
         Since the phase relations among the scattered 
         wavelets change with the scattering direction
         and the size and shape of the dust,
         the scattering of light by dust depends on the size, 
         shape, and chemical composition of the dust 
         and the direction at which the light is scattered. 
         }
\end{figure}

\vspace{-4mm}
\section{Scattering of Light by Dust: 
         A Conceptual Overview\label{sec:Phys_Scat}}
\vspace{-2mm}
When a dust grain, composed of discrete electric charges,
is illuminated by an electromagnetic wave,
the electric field of the incident electromagnetic wave 
will set the electric charges in the dust 
into oscillatory motion. These accelerated electric 
charges radiate electromagnetic energy in all directions,
at the same frequency as that of the incident wave.
This process, known as ``scattering'' 
(to be more precise, elastic scattering),
removes energy from the incident beam of 
electromagnetic radiation. 
Absorption also arises as the excited charges
transform part of the incident electromagnetic 
energy into thermal energy.
The combined effect of absorption and scattering,
known as ``extinction'', is the total energy loss of
the incident wave \cite{BH_1983}. 

The scattering of light by dust depends on the size, shape,
and chemical composition of the dust and the direction at
which the light is scattered. 
This can be qualitatively understood, 
as schematically shown in Figure \ref{fig:scat1},
by conceptually subdividing the dust into many small regions,
in each of which a dipole moment will be induced
when illuminated by an incident electromagnetic wave.
These dipoles oscillate at the frequency of the incident wave
and therefore scatter secondary radiation in all directions. 
The total scattered field of a given direction (e.g. $P_1$, $P_2$)
is the sum of the scattered wavelets, with their phase differences
taken into account. Since these phase relations change for 
a different scattering direction, the scattered field varies 
with scattering direction.\footnote{%
  An exception to this is the dust in the Rayleigh regime 
  (i.e. with its size being much smaller than the wavelength) 
  which scatters light nearly isotropically,
  with little variation with direction
  since for dust so small all the secondary wavelets 
  are approximately in phase.
  }
The phase relations among the scattered wavelets 
also change with the size and shape of the dust. 
On the other hand, the amplitude and phase of 
the induced dipole moment for a given frequency 
depend on the material of which the dust is composed. 
Therefore, the scattered field is sensitive to
both the size, shape and chemical composition of
the dust \cite{BH_1983}. 

\begin{figure}
\centering
\includegraphics[height=10cm]{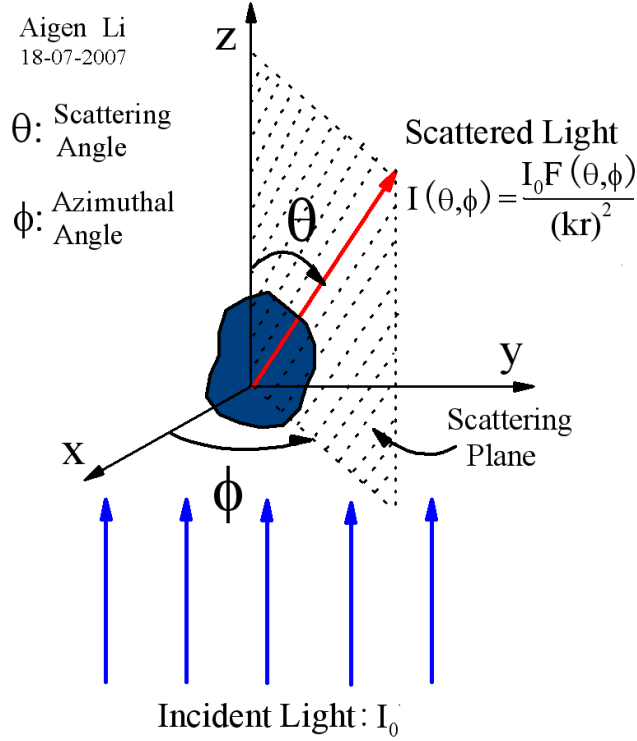}
%
\caption{Schematic scattering geometry of a dust grain 
         in an incident radiation field of intensity $\Io$
         which scatters radiation of intensity $I(\theta,\phi)$
         into a scattering angle $\theta$
         ($\theta$\,=\,$0^{\rm o}$: forward scattering;
          $\theta$\,=\,$180^{\rm o}$: backward scattering),
         an azimuthal angle $\phi$,
         and a distance $r$ from the dust. 
         In a Cartesian coordinate system, 
         the incident direction defines the $+$$z$ axis.
         The scattering direction and the incident direction
         define the scattering plane. 
         In the far-field region (i.e. $kr$\,$\gg$\,1),
         $I$\,=\,$\Io$\,$F(\theta,\phi)$/$k^2r^2$,
         where $k$\,=\,$2\pi/\lambda$ is the wave number in vacuum.    
         }
\label{fig:scat2}       
\end{figure}

\vspace{-2mm}
\section{Scattering of Light by Dust: Definitions\label{sec:Scat_Def}}
\vspace{-2mm}
As discussed in \S\ref{sec:Phys_Scat}, when light impinges 
on a grain it is either scattered or absorbed. 
Let $I_{\rm o}(\lambda)$ be the intensity of the incident 
light at wavelength $\lambda$,
the intensity of light scattered into a direction 
defined by $\theta$ and $\phi$ 
(see Fig.\,\ref{fig:scat2}) is
\begin{equation}
I(\lambda)\,=\,\frac{I_{\rm o}(\lambda)\,F(\theta,\phi)}{k^{2}\,r^{2}}
\end{equation}
where $0^{\rm o}$\,$\le$\,$\theta$\,$\le$\,$180^{\rm o}$
is the {\it scattering angle}
(the angle from the incident direction),
$0^{\rm o}$\,$\le$\,$\phi$\,$\le$\,$360^{\rm o}$
is the azimuthal angle
which uniquely determines the {\it scattering plane}
defined by the incident direction and the scattering 
direction (see Fig.\,\ref{fig:scat2}),\footnote{%
  When the scattering is along
  the incident direction ($\theta$\,=\,$0^{\rm o}$, 
  i.e. ``{\it forward scattering}'')
  or the scattering is on the opposite direction of
  the incident direction ($\theta$\,=\,$180^{\rm o}$, 
  i.e. ``{\it backward scattering}''),
  any plane containing the $z$ axis is a suitable scattering plane.
  }
$F(\theta,\,\phi)$ is the (dimensionless) angular scattering function,
$r$\,$\gg$\,$\lambda/2\pi$ is the distance from the scatterer, 
and $k$\,=\,$2\pi/\lambda$ is the wave number in vacuum.
The {\it scattering cross section} $\Csca$, 
defined as the area on which the incident wave falls 
with the same amount of energy as that scattered 
in all directions by the dust,
may be obtained by integrating 
the angular scattering distribution 
$F(\theta,\phi)/k^2$ over all solid angles
\begin{equation}
\Csca(\lambda)\,=\,\frac{1}{k^2}\int_{0}^{2\pi}\int_{0}^{\pi} 
              F(\theta,\phi)\,\sin\theta\,d\theta\,d\phi ~~~,
\end{equation}
where $F(\theta,\phi)/k^2$ (with a dimension of area),\footnote{%
  Also called
  the ``{\it differential scattering cross section}''
  $d\Csca/d\Omega$\,$\equiv$\,$F(\theta,\phi)/k^2$,
  it specifies the angular distribution of the scattered light
  [i.e. the amount of light (for unit incident irradiance)
  scattered into a unit solid angle about a given direction].
  }
after normalized by $\Csca$, is known as the {\it phase function}
or {\it scattering diagram}
%
\begin{equation}
  p(\theta,\phi)\,\equiv\,\frac{F(\theta,\phi)/k^2}{\Csca} ~~. 
\end{equation}
The {\it asymmetry parameter} (or {\it asymmetry factor}) $g$ is
defined as the average cosine of the scattering angle $\theta$
\begin{equation}
  g \equiv \langle\cos\theta\rangle = \int_{4\pi} 
  p(\theta,\phi)\,\cos\theta\,d\Omega
  = \frac{1}{k^2\,\Csca}
    \int_{0}^{2\pi}\int_{0}^{\pi} 
    F(\theta,\phi)\,\cos\theta\,\sin\theta\,d\theta\,d\phi ~~~.
\end{equation}
The {\it asymmetry parameter} $g$,
specifying the degree of scattering in 
the forward direction ($\theta$\,=\,0$^{\rm o}$),
varies from $-1$ (i.e. all radiation is backward 
scattered like a ``mirror'') 
to $1$ (for pure forward scattering).
If a grain scatters more light toward
the forward direction, $g$\,$>$\,0;
$g$\,$<$\,0 if the scattering is directed
more toward the back direction;
$g$\,=\,0 if it scatters light isotropically
(e.g. small grains in the Rayleigh regime)
or if the scattering is symmetric with respect to
$\theta$\,=\,90$^{\rm o}$
(i.e. the scattered radiation is azimuthal
independent and symmetric with respect to
the plane perpendicular to the incident radiation).

In radiative transfer modeling of dusty regions, 
astronomers often use the empirical
Henyey-Greenstein phase function to represent 
the anisotropic scattering properties of dust \cite{HG_PhaseFunc1941}
\begin{equation}\label{eq:HG1941}
H(\theta)\,\equiv\,\frac{1}{4\pi} 
  \frac{1-g^2}
  {\left(1+g^2-2g\cos\theta\right)^{3/2}} ~~.
\end{equation}
Draine (2003) proposed a more general analytic form
for the phase function
\begin{equation}\label{eq:Draine2003}
H_\eta(\theta)\,\equiv\,\frac{1}{4\pi}\, 
  \frac{3\left(1-g^2\right)}
  {3+\eta \left(1+2g^2\right)}\,
  \frac{1+\eta \cos^2\theta}
  {\left(1+g^2-2g\cos\theta\right)^{3/2}} 
\end{equation}
where $\eta$ is an adjustable parameter.
For $\eta$\,=\,0 Eq.\ref{eq:Draine2003}
reduces to the Henyey-Greenstein phase function. 
For $g$\,=\,0 and $\eta$\,=\,1 
Eq.\ref{eq:Draine2003} reduces to the phase function 
for Rayleigh scattering \cite{Draine_UV_Scat_2003}.
 
As discussed in \S\ref{sec:Phys_Scat}, both scattering
and absorption (the sum of which is called {\it extinction}) 
remove energy from the incident beam.
The {\it extinction cross section}, 
defined as
\begin{equation}
\Cext = \Csca + \Cabs
 = \frac{\rm\small total\,energy\,scattered\,
           and\,absorbed\,per\,unit\,time}
          {\rm\small incident\,energy\,
           per\,unit\,area\,per\,unit\,time} ~,
\end{equation}
is determined from the {\it optical theorem} 
which relates $\Cext$ to the real part of 
the complex scattering amplitude 
$S(\theta,\phi)$\footnote{%
  The angular scattering function $F(\theta,\phi)$
  is just the absolute square of the complex
  scattering amplitude $S(\theta,\phi)$:
  $F(\theta,\phi) = |S(\theta,\phi)|^2$.
  }
in the forward direction alone \cite{Jackson_EM}
\begin{equation}
\Cext = -\frac{4\pi}{k^2}\,{\rm Re}\{S(\theta=0^{\rm o})\} ~~.
\end{equation}
The {\it absorption cross section} $\Cabs$ is 
the area on which the incident wave falls with the same 
amount of energy as that absorbed inside the dust;
$\Cext$, having a dimension of area, 
is the ``effective'' blocking area to the incident radiation
(for grains much larger than the wavelength of the incident
radiation, $\Cext$ is about twice the geometrical blocking area). 
For a grain (of size $a$ and complex index of refraction $m$)
in the Rayleigh limit 
(i.e. $2\pi a/\lambda$\,$\ll$\,1,
      $2\pi a |m|/\lambda$\,$\ll$\,1),
the absorption cross section $\Cabs$ is much larger than
the scattering cross section $\Csca$ 
and therefore $\Cext \approx \Cabs$.
Non-absorbing dust has $\Cext=\Csca$.

The {\it albedo} of a grain is defined as
$\alpha$\,$\equiv$\,$\Csca/\Cext$.
For grains in the Rayleigh limit,
$\alpha$\,$\approx$\,0 since $\Csca$\,$\ll$\,$\Cabs$.
For Non-absorbing dust, $\alpha$\,=\,1.

In addition to energy, light carries momentum 
of which the direction is that of propagation 
and the amount is $h\nu/c$ (where $h$ is the Planck
constant, $c$ is the speed of light, 
and $\nu$ is the frequency of the light).
Therefore, upon illuminated by an incident beam of light,
dust will acquire momentum and a force called 
{\it radiation pressure} will be exerted on it
in the direction of propagation of the incident light.
The radiation pressure force is proportional to
the net loss of the forward component of the momentum 
of the incident beam. 
While the momentum of the {\it absorbed} light 
(which is in the forward direction)
will all be transfered to the dust,
the forward component of the momentum 
of the {\it scattered} light will not
be removed from the incident beam.
Therefore, the radiation pressure force
exerted on the dust is
\begin{equation}
F_{\rm pr} = \Io \Cpr/c ~~,~~ 
\Cpr = \Cabs + \left(1-g\right)\,\Csca ~,
\end{equation}
where $\Cpr$ is the radiation pressure cross section, 
and $\Io$ is the intensity (irradiance) of the incident light.

In literature, one often encounters 
$\Qext$, $\Qsca$, $\Qabs$, and $\Qpr$ 
-- the extinction, scattering, absorption 
and radiation pressure efficiencies.
They are defined as the extinction, scattering, 
absorption, and radiation pressure cross sections 
divided by the geometrical cross section of 
the dust $\Cgeo$,
\begin{equation}
\Qext = \frac{\Cext}{\Cgeo} ~;~~
\Qsca = \frac{\Csca}{\Cgeo} ~;~~
\Qabs = \frac{\Cabs}{\Cgeo} ~;~~
\Qpr = \frac{\Cpr}{\Cgeo} ~.
\end{equation}
For spherical grains of radii $a$, 
$\Cgeo$\,=\,$\pi a^2$.
For non-spherical grains, there is no uniformity
in choosing $\Cgeo$. 
A reasonable choice is the geometrical cross section
of an ``equal volume sphere'' 
$\Cgeo$\,$\equiv$\,$\pi\left(3V/4\pi\right)^{2/3}$\,
$\approx$\,1.21\,$V^{2/3}$ where $V$ is the volume
of the non-spherical dust.

\section{Scattering of Light by Dust: 
         Dielectric Functions
         \label{sec:Diel_Func}}
\vspace{-2mm}
The light scattering properties of dust are usually 
evaluated based on its optical properties
(i.e. {\it dielectric functions} or 
{\it indices of refraction})  
and geometry (i.e. size and shape) 
by solving the Maxwell equations
\begin{equation}
\nabla \times \bE + \frac{1}{c}\frac{\partial{\bB}}{\partial t} = 0 ~;
\nabla \times \bH - \frac{1}{c}\frac{\partial{\bD}}{\partial t}
= \frac{4\pi}{c} \bJ ~;
\nabla\cdot \bD = 4\pi \rho ~;
\nabla\cdot \bB = 0 ~;
\end{equation}
where $\bE$ is the electric field,
$\bB$ is the magnetic induction,
$\bH$ is the magnetic field,
$\bD$ is the electric displacement,
$\bJ$ is the electric current density,
and $\rho$ is the electric charge density.
They are supplemented with 
the {\it constitutive relations} (or ``{\it material relations}'')
\begin{equation}
\bJ = \sigma\bE ~;~ 
\bD = \varepsilon \bE = \left(1+4\pi\chi\right) \bE~;~
\bB = \mu \bH ~;
\end{equation}
where $\sigma$ is the {\it electric conductivity},
$\varepsilon$\,=\,$1+4\pi\chi$ is the {\it dielectric function} 
(or {\it permittivity} or {\bf dielectric permeability}), 
$\chi$ is the {\it electric susceptibility},
and $\mu$ is the {\it magnetic permeability}.
For time-harmonic fields 
$\bE$,\,$\bH$\,$\propto$\,$\exp\left(-i\,\omega t\right)$, 
the Maxwell equations are reduced to the Helmholtz wave equations
\begin{equation}
\nabla^2\bE + \tilde{k}^2\bE =0 ~;~
\nabla^2\bH + \tilde{k}^2\bH =0 ~;
\end{equation}
where $\tilde{k} = m\,\omega/c$ is the complex wavenumber,
$m = \sqrt{\mu\left(\varepsilon + i\,4\pi\sigma/\omega\right)}$
is the complex refractive index,
and $\omega$\,=\,$2\pi c/\lambda$ is the circular frequency.
These equations should be considered for the field
outside the dust (which is the superposition of
the incident field and the scattered field)
and the field inside the dust,
together with the boundary conditions
(i.e. any tangential and normal components 
of $\bE$ are continuous across the dust boundary).
For non-magnetic dust ($\mu$\,=\,1),
the complex refractive index is
$m = \sqrt{\varepsilon + i\,4\pi\sigma/\omega}$.
For non-magnetic, non-conducting dust,
$m = \sqrt{\varepsilon}$.
For highly-conducting dust,
$4\pi\sigma/\omega$\,$\gg$\,1,
therefore, both the real part and the imaginary
part of the index of refraction are approximately
$\sqrt{2\pi\sigma/\omega}$
at long wavelengths.

The complex refractive indices $m$ 
or dielectric functions $\varepsilon$
of dust are often called {\it optical constants}, 
although they are not constant but vary with wavelengths.
They are of great importance in studying the absorption
and scattering of light by dust.
They are written in the form of $m = \mre + i\,\mim$
and $\varepsilon = \epsilonre + i\,\epsilonim$.
The sign of the imaginary part of $m$ or $\varepsilon$ 
is opposite to that of 
the time-dependent term of the harmonically variable fields
[i.e. $\mim$, $\epsilonim$\,\,$>$\,0 
for $\bE$,\,$\bH$\,$\propto$\,$\exp\left(-i\,\omega t\right)$; 
$\mim$, $\epsilonim$\,\,$<$\,0 
for $\bE$,\,$\bH$\,$\propto$\,$\exp\left(i\,\omega t\right)$]. 
The imaginary part of the index of refraction
characterizes the attenuation of the wave
($4\pi\mim/\lambda$ is called the {\it absorption coefficient}),
while the real part determines the {\it phase velocity}
($c/\mre$) of the wave in the medium:
for an electric field propagating
in an absorbing medium of $m = \mre + i\,\mim$,
say, in the $x$ direction, 
$\bE \propto \exp(-\omega \mim x/c) \exp[-i\omega(t-\mre x/c)]$.

The physical basis of the dielectric function $\varepsilon$
can readily be understood in terms of the 
classical {\it Lorentz harmonic oscillator} model
(for insulators) and the {\it Drude} model
(for free-electron metals). 
In the Lorentz oscillator model, the {\it bound} electrons and ions of 
a dust grain are treated as simple harmonic oscillators 
subject to the driving force of an applied electromagnetic field.
The applied field distorts the charge distribution and 
therefore produces an {\it induced dipole moment}.
To estimate the induced moments we consider 
a (polarizable) grain as a collection of 
identical, independent, isotropic, 
harmonic oscillators with mass $m$ and charge $q$; 
each oscillator is under the action of three forces:
(i) a restoring force $-K\bx$, 
    where $K$ is the force constant (i.e. stiffness)
    of the ``spring'' to which the bound charges are attached, 
    and $\bx$ is the displacement of the bound charges from 
    their equilibrium;
(ii) a damping force $-b\dotx$,
     where $b$ is the damping constant; and
(iii) a driving force $q\bE$ 
      produced by the (local) electric field $\bE$.
The equation of motion of the oscillators is
\begin{equation}\label{eq:motion_bound}
m\ddotx + b\dotx + K\bx = q\bE ~~.
\end{equation}
For time harmonic electric fields 
$\bE \propto \exp\left(-i\omega t\right)$, 
we solve for the displacement 
\begin{equation}
\bx = \frac{\left(q/m\right)\bE}{\omegao^2-\omega^2-i\gamma\omega} ~,
~ \omegao^2 = K/m~, ~\gamma=b/m ~,
\end{equation}
where $\omegao$ is the frequency of oscillation about equilibrium
(i.e. the restoring force is $-m\omegao^2\bx$).
The induced dipole moment $\dipmom$ of an oscillator
is $\dipmom=q\bx$.
Let $n$ be the number of oscillators per unit volume.
The {\it polarization} (i.e. dipole moment per unit 
volume) $\pol = n\dipmom=n q\bx$ is
\beq
\pol = \frac{1}{4\pi} 
       \frac{\omegap^2}{\omegao^2-\omega^2-i\gamma\omega}\bE ~~,
~~ \omegap^2=4\pi n q^2/m ~, 
\eeq  
where $\omegap$ is the plasma frequency.
Since $\pol = \chi\bE = \left(\varepsilon - 1\right)/4\pi\,\bE$,
the dielectric function for a one-oscillator model
around a resonance frequency $\omegao$ is 
\beq\label{eq:epsilon_Lorentz}
\varepsilon = 1 + 4\pi\chi 
= 1 + \frac{\omegap^2}{\omegao^2-\omega^2-i\gamma\omega}
\eeq 
\beq\label{eq:epsilon_Lorentz_Re}
\epsilonre = 1 + 4\pi\chi^{\prime} 
= 1 + \frac{\omegap^2\left(\omegao^2-\omega^2\right)}
{\left(\omegao^2-\omega^2\right)^2+\gamma^2\omega^2}
\eeq 
\beq\label{eq:epsilon_Lorentz_Im}
\epsilonim = 4\pi\chi^{\prime\prime} 
= \frac{\omegap^2\gamma\omega}
{\left(\omegao^2-\omega^2\right)^2+\gamma^2\omega^2} ~~~.
\eeq
We see from Eq.\ref{eq:epsilon_Lorentz_Im} $\epsilonim$\,$\ge$\,0 
for all frequencies. This is true for materials close to
thermodynamic equilibrium, except those with population inversions.
If there are many oscillators of different frequencies,
the dielectric function for a multiple-oscillator model is
\beq
\varepsilon 
= 1 + \sum_j\frac{\omega_{{\rm p},j}^2}{\omega_j^2-\omega^2-i\gamma_j\omega}
\eeq 
where $\omega_{{\rm p},j}$, $\omega_j$ and $\gamma_j$
are respectively the plasma frequency, the resonant frequency,
and the damping constant of the $j$-th oscillator.

\begin{figure}
\centering
\includegraphics[height=10cm]{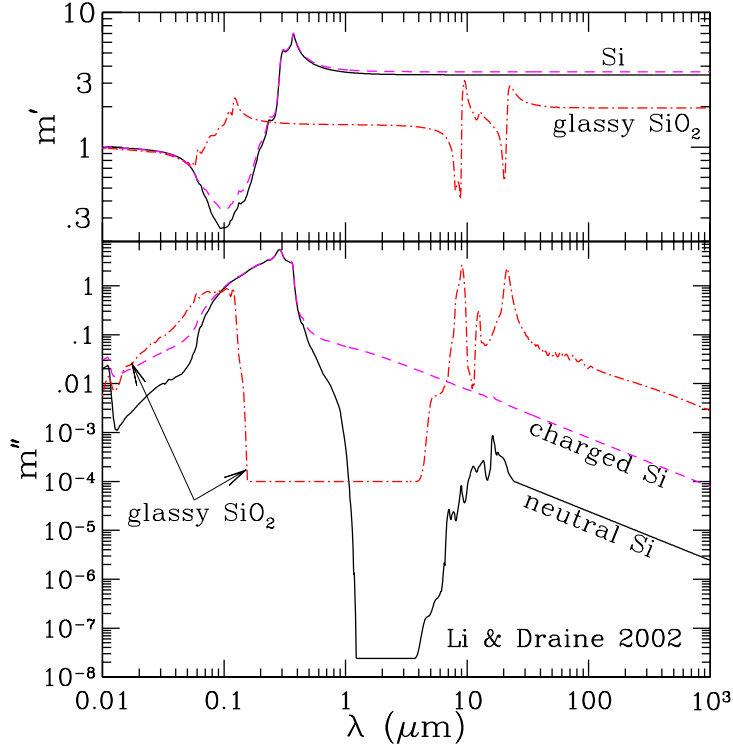}
%
\caption{Refractive indices $\mre$ (upper panel), $\mim$ (lower panel) 
        of neutral silicon nanoparticles (Si; solid lines), 
        singly, positively charged ($Z=+1$) silicon nanoparticles
        of size $a=10\Angstrom$ (dashed lines), 
        and SiO$_2$ glass (dot-dashed lines). 
        Compared to neutral Si,
        in charged Si free electrons (if negative charged)
        or holes (if positively charged) contribute
        to the dielectric function. 
        Crystalline Si is IR inactive since its lattice vibrations 
        have no dipole moment; the  bands at 6.91, 7.03, 
        7.68, 8.9, 11.2, 13.5, 14.5, 16.4, and 17.9$\mu$m 
        are due to multi-phonon processes.
        Taken from \cite{Li_Draine_2002_silicon}.
         }
\label{fig:si_nk}       
\end{figure}

The optical properties associated with {\it free electrons}
are described by the {\it Drude} model. The free electrons
experience no restoring forces when driven by the electric 
field of a light wave and do not have natural resonant 
frequencies (i.e. $\omegao$\,=\,0). 
The Drude model for metals is obtained 
directly from the Lorentz model for insulators simply by
setting the restoring force in Eq.\ref{eq:motion_bound}
equal to zero. The dielectric function for free electrons is
\beq\label{eq:epsilon_Drude}
\varepsilon  
= 1 - \frac{\omegape^2}{\omega^2+i\gamma_e\omega} ~~,~~
\omegape^2=4\pi n_e e^2/m_e 
\eeq 
\beq\label{eq:epsilon_Drude_Re}
\epsilonre 
= 1 - \frac{\omegape^2}{\omega^2+\gamma_e^2}
\eeq 
\beq\label{eq:epsilon_Drude_Im}
\epsilonim = \frac{\omegape^2\gamma_e}
{\omega\left(\omega^2+\gamma_e^2\right)} ~~~
\eeq
where $\omegape$ is the plasma frequency,
$n_e$ is the density of free electrons 
and $m_e$ is the effective mass of an electron.
The damping constant $\gamma_e$\,=\,$1/\tau_e$
is the reciprocal of the mean free time between 
collisions ($\tau_e$) which are often determined by
electron-phonon scattering (i.e. interaction 
of the electrons with lattice vibrations).\footnote{%
  For nano-sized metallic dust of size $a$ 
  (which is smaller than the mean free path of 
  conduction electrons in the bulk metal), 
  $\tau_e$ and $\gamma_e$ are 
  increased because of additional collisions with
  the boundary of the dust: 
  $\gamma_e = \gamma_{\rm bulk} + v_F/(\varsigma a)$, 
  where $\gamma_{\rm bulk}$ is the bulk metal damping constant, 
  $v_F$ is the electron velocity at the Fermi surface, 
  $\varsigma$ is a dimensionless constant of order unity which 
  depends on the character of the scattering at the boundary
  ($\varsigma a$ is the effective mean free path for collisions
   with the boundary): 
  $\varsigma$\,=\,1 for classic isotropic scattering, 
  $\varsigma$\,=\,4/3 for classic diffusive scattering, 
  $\varsigma$\,=\,1.16 or $\varsigma$\,=\,1.33 for scattering 
  based on the quantum particle-in-a-box model 
  (see \cite{Coronado_Schatz_2003} and
  references therein). 
  Since $\omega^2$\,$\gg$\,$\gamma_e^2$ in metals
  near the plasma frequency, $\epsilonim$ can be written as 
  $\epsilonim$\,=\,$\epsilonim_{\rm bulk} + 
  v_F \omegap^2/(\varsigma a \omega^3)$.
  This, known as the ``electron mean free path limitation'' effect,
  indicates that for a metallic grain 
  $\epsilonim$ increases as the grain becomes smaller.
  }
For dust materials (e.g. graphite) containing
both bound charges and free electrons, 
the dielectric function is 
\beq\label{eq:epsilon_both}
\varepsilon = 1 - \frac{\omegape^2}{\omega^2+i\gamma_e\omega}
+ \sum_j\frac{\omega_{{\rm p},j}^2}
             {\omega_j^2-\omega^2-i\gamma_j\omega} ~~.
\eeq 

The optical response of free electrons in metals
can also be understood in terms of the electric current
density $\bJ$ and conductivity $\sigma(\omega)$.
The free electrons in metals move between molecules.
In the absence of an external field, they move in 
a random manner and hence they do not give rise to
a net current flow. When an external field is applied,
the free electrons acquire an additional velocity
and their motion becomes more orderly which gives
rise to an induced current flow.  
The current density $\bJ=-n_e e\dotx$ is obtained
by solving the equation of motion of the free electrons 
$m_e \ddotx = -e\bE - m_e \dotx/\tau_e$
\beq
\dotx = \frac{-\tau_e e}{m_e}
        \frac{1}{1 - i\,\omega\tau_e}\bE ~~.
\eeq
Since $\bJ=\sigma\bE$, the a.c. conductivity
is $\sigma = \sigma_{\rm o}/\left(1-i\,\omega\tau_e\right)$,
where $\sigma_{\rm o} = n_e\tau_e e^2/m_e$ is
the d.c. conductivity. The dielectric function 
for a free-electron metal is therefore
\beq
\varepsilon = 1 + \frac{i\,4\pi\sigma}{\omega}
= 1 - \frac{\omegape^2}{\omega^2+i\omega/\tau_e} ~~.
\eeq

The real and imaginary parts of the dielectric function
are not independent. They are related through 
the {\it Kramers-Kronig} (or {\it dispersion}) relations   
\begin{equation}\label{eq:kk_epsilon}
\epsilonre(\omega) = 1 + \frac{2}{\pi}P\int_{0}^{\infty}
\frac{x\epsilonim(x)}{x^2-\omega^2}\,dx ~~,~~
\epsilonim(\omega) = \frac{-2\omega}{\pi}P\int_{0}^{\infty}
\frac{\epsilonre(x)}{x^2-\omega^2}\,dx ~~,
\end{equation}
where $P$ is the Cauchy Principal value of the integral
\begin{equation}\label{eq:cauchy}
P\int_{0}^{\infty} \frac{x\epsilonim(x)}{x^2-\omega^2}\,dx
= \lim\limits_{a\rightarrow 0} \left[\int_{0}^{\omega-a}
\frac{x\epsilonim(x)}{x^2-\omega^2}\,dx
+ \int_{\omega+a}^{\infty}
\frac{x\epsilonim(x)}{x^2-\omega^2}\,dx\right] ~~.
\end{equation}

The real and imaginary parts of the
index of refraction are also connected through
the Kramers-Kronig relation.
This also holds for the real and imaginary parts of 
the electric susceptibility.
\begin{equation}\label{eq:kk_nk1}
\mre(\omega) = 1 + \frac{2}{\pi}P\int_{0}^{\infty}
\frac{x\mim(x)}{x^2-\omega^2}\,dx ~~,~~
\mim(\omega) = \frac{-2\omega}{\pi}P\int_{0}^{\infty}
\frac{\mre(x)}{x^2-\omega^2}\,dx ~~,
\end{equation}
\begin{equation}\label{eq:kk_chi}
\chi^{\prime}(\omega) = \frac{2}{\pi}P\int_{0}^{\infty}
\frac{x\chi^{\prime\prime}(x)}{x^2-\omega^2}\,dx ~~,~~
\chi^{\prime\prime}(\omega) = \frac{-2\omega}{\pi}P\int_{0}^{\infty}
\frac{\chi^{\prime}(x)}{x^2-\omega^2}\,dx ~~.
\end{equation}

The Kramers-Kronig relation can be used to relate 
the extinction cross section integrated over 
the entire wavelength range to the dust volume $V$
\begin{equation}\label{eq:Purcell_KK}
\int_{0}^{\infty} \cext(\lambda)\,d\lambda 
= 3 \pi^2 F V~~~,
\end{equation}
where $F$, a dimensionless factor, 
is the orientationally-averaged polarizability 
relative to the polarizability of an equal-volume 
conducting sphere, depending only upon the grain shape 
and the static (zero-frequency) dielectric constant 
$\epsilon_{\rm o}$ of the grain material 
\cite{Purcell_1969}.
This has also been used to place constraints
on interstellar grain models based on the interstellar 
depletions \cite{Li_2005,Li_2006}
and the carrier of the mysterious 21$\mum$ emission
feature seen in 12 protoplanetary nebulae \cite{Li_TiC_2003}.

For illustration, we show in Figure \ref{fig:si_nk}
the refractive indices of glassy SiO$_2$ and neutral
and singly-charged silicon nanoparticles (SNPs).\footnote{%
  SNPs were proposed as the carrier of the ``extended red emission''
  (ERE), a broad, featureless emission band between
  $\simali$5400 and 9000$\Angstrom$ seen in a wide
  variety of dusty environments 
  \cite{Witt_etal1998,Ledoux_etal1998}
  (but see \cite{Li_Draine_2002_silicon}).
  SNPs were also suggested to be present in
  the inner corona of the Sun \cite{Habbal_etal2003}
  (but see \cite{Mann_nano_2007,Mann_Murad2005}).  
  }
The optical properties of SNPs depend on
whether any free electrons or holes are present.
The contribution of free electrons or holes 
to the dielectric function of SNPs is approximated by
$\delta\varepsilon \approx -\omegap^2\tau^2/
\left(\omega^2\tau^2 + i\omega\tau\right)$,
$\omegap^2 = 3|Z|\,e^2/a^3 m_{\rm eff}$,
where $e$ is the proton charge,
$Ze$ is the grain charge, 
$a$ is the grain radius,
$\tau\approx a/v_F$ is the collision time
(we take $v_F$\,$\approx$\,$10^8\cm\s^{-1}$),
and $m_{\rm eff}$ is the effective mass of 
a free electron or hole.
In Figure \ref{fig:si_nk} the charged SNPs are taken
to contain only one hole (i.e. $Z$\,=\,$+1$). 
We take $m_{\rm eff}$\,$\approx$\,$0.2\,m_e$.

\vspace{-4mm}
\section{Scattering of Light by Dust: Analytic Solutions
         \label{sec:Scat_Approx}}
\vspace{-2mm}
For dust with sizes much smaller than the wavelength
of the incident radiation, analytic solutions to the
light scattering problem exist for certain shapes.
Let $a$ be the characteristic length of the dust,
and $x$\,$\equiv$\,$2\pi a/\lambda$ be the dimensionless
{\it size parameter}.
Under the condition of $x$\,$\ll$\,1 and $|mx|$\,$\ll$\,1 
(i.e. in the ``{\it Rayleigh}'' regime), 
$\cabs$\,=\,$4\pi k \,{\rm Im}\{\balpha\}$,
$\csca$\,=\,$\left(8\pi/3\right)\, k^4 \,|\balpha|^2$,
where $\balpha$ is the complex electric {\it polarizability}
of the dust. Apparently, $\csca$\,$\ll$\,$\Cabs$
and $\Cext$\,$\approx$\,$\Cabs$.
In general, $\balpha$ is a diagonalized tensor;\footnote{%
  $\balpha$ can be diagonalized by appropriate choice
  of Cartesian coordinate system. It describes the linear
  response of a dust grain to applied electric field $\bE$:
  $\dipmom$\,=\,$\balpha\bE$ where $\dipmom$ is 
  the induced electric dipole moment.}
for homogeneous spheres 
composed of an isotropic material,
it is independent of direction 
\begin{equation}\label{eq:pol_spheres}
\alpha_{jj} = \frac{3V}{4\pi} \frac{\varepsilon -1}
{\varepsilon + 2}  ~~,
\end{equation}
where $V$ is the dust volume.\footnote{%
  For a dielectric sphere with dielectric function
  given in Eq.(\ref{eq:epsilon_Lorentz}),
  in the Rayleigh regime the absorption cross section is
  \beq
  \cabs(\omega)/V = \frac{9}{c}
  \frac{\gamma\omegap^2\,\omega^2}
       {\left(3\omega^2-\omegap^2-3\omegao^2\right)^2
              + 9\gamma^2\omega^2} ~~.
  \eeq
  Similarly, for a metallic sphere with dielectric function
  given in Eq.(\ref{eq:epsilon_Drude}),
  \beq
  \cabs(\omega)/V = \frac{9}{c}
  \frac{\gamma_e\omegape^2\,\omega^2}
       {\left(3\omega^2-\omegape^2\right)^2
              + 9\gamma_e^2\omega^2} ~~.
  \eeq
  It is seen that the frequency-dependent 
  absorption cross section for both dielectric
  and metallic spheres is a Drude function. 
  This is also true for ellipsoids.
  }
For a homogeneous, isotropic ellipsoid,
the polarizability for electric field vector
parallel to its principal axis $j$ is
\begin{equation}\label{eq:pol_spheroids}
\alpha_{jj} = \frac{V}{4\pi} 
\frac{\varepsilon-1}{(\varepsilon-1)L_j + 1} ~~,
\end{equation}
where $L_j$ is the ``depolarization factor'' 
along principal axis $j$ (see \cite{Draine_Lee_1984}).
The electric polarizability $\balpha$ is also known
for concentric core-mantle spheres \cite{vandehulst_1957},
confocal core-mantle ellipsoids
\cite{Draine_Lee_1984,Gilra_1972},
and multi-layered ellipsoids of {\it equal-eccentricity}
\cite{Farafonov_2001}.
For a thin {\it conducting} cylindrical rod
with length $2l$ and radius $\ahat$\,$\ll$\,$l$,
the polarizability along the axis of the rod is
\cite{Landau_Lifshitz_Pitaevskii_2000}
\begin{equation}
\alpha_{jj} \approx \frac{l^3}
{3\log\left(4l/\ahat\right) -7} ~~.
\end{equation}

In astronomical modeling, 
the most commonly invoked 
grain shapes are spheres 
and spheroids (oblates or prolates).\footnote{%
  Spheroids are a special class of ellipsoids.
  Let $\ahat$, $\bhat$, and $\chat$ be the semi-axes of
  an ellipsoid. For spheroids, $\bhat$\,=\,$\chat$.
  Prolates with $\ahat$\,$>$\,$\bhat$ are generated 
  by rotating an ellipse (of semi-major axis $\ahat$   
  and semi-minor axis $\bhat$) about its major axis;
  oblates with $\ahat$\,$<$\,$\bhat$ are generated 
  by rotating an ellipse (of semi-minor axis $\ahat$   
  and semi-major axis $\bhat$) about its minor axis. 
  }
In the Rayleigh regime, their absorption and scattering
properties are readily obtained from 
Eqs.(\ref{eq:pol_spheres},\ref{eq:pol_spheroids}).
For both dielectric and conducting spheres
(as long as $x$\,$\ll$\,1 and $|mx|$\,$\ll$\,1) 
\beq\label{eq:rayleigh_spheres}
\cabs/V = \frac{9\,\omega}{c} 
\frac{\epsilonim}{\left(\epsilonre+2\right)^2+\epsilonim^2} ~~\gg~~
\Csca/V = \frac{3}{2\pi}\left(\frac{\omega}{c}\right)^4
\left|\frac{\varepsilon-1}{\varepsilon+2}\right|^2 ~~.
\eeq 
At long wavelengths, for dielectric dust
$\epsilonim \propto \omega$ while $\epsilonre$ approaches 
a constant much larger than $\epsilonim$
(see Eqs.\ref{eq:epsilon_Lorentz_Re},\ref{eq:epsilon_Lorentz_Im}),
we see $\Cabs\propto \omega\,\epsilonim \propto \omega^2$; 
for metallic dust, $\epsilonim \propto 1/\omega$
while $\epsilonre$ approaches a constant much smaller
than $\epsilonim$
(see Eqs.\ref{eq:epsilon_Drude_Re},\ref{eq:epsilon_Drude_Im}),
we see $\cabs\propto \omega/\epsilonim \propto \omega^2$; 
therefore, for both dielectric and metallic dust 
$\cabs \propto \lambda^{-2}$ at long wavelengths!\footnote{%
   However, various astronomical data suggest a {\it flatter} 
   wavelength-dependence
   (i.e. $\cabs$\,$\propto$\,$\lambda^{-\beta}$ 
    with $\beta$\,$<$\,2):
    $\beta$$<$2 in the far-IR/submm wavelength range
    has been reported for interstellar molecular clouds, 
    circumstellar disks around young stars, and
    circumstellar envelopes around evolved stars.
    Laboratory measurements have also found $\beta$\,$<$\,2 
    for certain cosmic dust analogues.
    In literature, the {\it flatter} ($\beta$$<$2) 
    long-wavelength opacity law is commonly attributed to 
    grain growth by coagulation of small dust into large 
    fluffy aggregates (see \cite{Li_AIP_2005} and references therein).
    However, as shown in Eq.(\ref{eq:Purcell_KK}), 
    the Kramers-Kronig relation requires
    that $\beta$ should be larger than 1 for 
    $\lambda$\,$\rightarrow$$\infty$ 
    since $F$ is a finite number 
    and the integration in the left-hand-side of 
    Eq.(\ref{eq:Purcell_KK}) 
    should be convergent although we cannot rule out 
    $\beta$\,$<$\,1 over a finite range of wavelengths. 
    }

It is also seen from Eq.(\ref{eq:rayleigh_spheres}) that
for spherical dust in the Rayleigh regime 
the albedo $\alpha\approx 0$,
and the radiation cross section 
$\Cpr\approx \Cabs$.
This has an interesting implication.
Let $\beta_{\rm pr}(a)$ be the ratio of 
the radiation pressure force to the gravity 
of a spherical grain of radius $a$ 
in the solar system or in debris disks
illuminated by stars (of radius $R_\star$
and mass $M_\star$) with a stellar flux of 
$F_\lambda^\star$ at the top of the atmosphere
\cite{Mann_COE_2007},
\begin{equation}\label{eq:betapr}
\beta_{\rm pr}(a) = \frac{3\,R_\star^2 
                \int F_\lambda^{\star} 
                \left[ C_{\rm abs}(a,\lambda) 
                + \left(1- g\right)
                   C_{\rm sca}(a,\lambda)\right] d\lambda}
                 {16\pi\,c\,G M_\star a^3 \rho_{\rm dust}}
\end{equation}
where $G$ is the gravitational constant,
and $\rho_{\rm dust}$ is the mass density
of the dust. For grains in the Rayleigh regime
($g$\,$\approx$\,0; $\csca$\,$\ll$\,$\cabs$;
$\cabs$\,$\propto$\,$a^3$),   
we see $\beta_{\rm pr}$\,$\propto$\,$\Cabs/a^3$ 
is independent of the grain size $a$ 
(see Fig.\,\ref{fig:beta_pr})! 

\begin{figure}
\centering
\includegraphics[height=8cm]{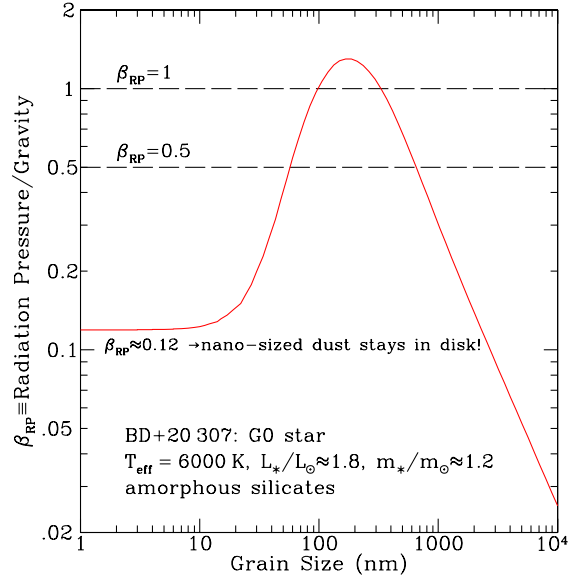}
\vspace{-3mm}
\caption{\label{fig:beta_pr}
         $\beta_{\rm pr}$ -- Ratio of radiation pressure to
         gravity for 
         compact silicate grains in the debris disk
         around the Sun-like star BD+20\,307 
         (G0, age $\sim$\,300\,Myr).   
         We note for nano-sized grains
         $\beta_{\rm pr}$\,$\approx$\,0.12,
         independent of grain size;
         for grains larger than $\simali$0.3$\mum$,
         $\beta_{\rm pr}$ is inverse proportional to
         grain size. Song et al. \cite{Song_etal_2005}
         attributed the dust in this disk 
         to recent extreme collisions between asteroids. 
         Taken from \cite{Li_Orteg_Lunine_2008}.
         }
\label{fig:beta_pr}
\end{figure}

Spheroids are often invoked to model the interstellar polarization.
In the Rayleigh approximation, their absorption cross sections
for light polarized parallel ($\|$) or perpendicular ($\bot$) 
to the grain symmetry axis are\footnote{%
  For grains spinning around the principal axis 
  of the largest moment of inertia,
  the polarization cross sections are 
  $\cppol = (\cpar-\cper)/2$ for prolates, 
  and $\copol = (\cper-\cpar)$ for oblates;
  the absorption cross sections for randomly-oriented spheroids are
  $\cabs = (\cpar+2\cper)/3$ \cite{Lee_Draine_1985}. 
  }
\beq\label{eq:rayleigh_spheroids}
C_{\rm abs}^{\|, \bot}/V = \frac{\omega}{c} {\rm Im}\left\{
            \frac{\varepsilon-1}{(\varepsilon-1)L_{\|,\bot}+1}\right\} ~,
\eeq 
where the depolarization factors parallel ($L_{\|}$) 
or perpendicular ($L_{\bot}$) 
to the grain symmetry axis are not independent,
but related to each other through
$L_{\|}$\,$+$\,2\,$L_{\bot}$\,=\,1, with
\begin{equation}
L_{\|} = \frac{1-\ehat^2}{\ehat^2} 
      \left[\frac{1}{2\ehat} {\rm ln}\left(\frac{1+\ehat}
      {1-\ehat}\right)-1\right] ~~,~~
\ehat = \sqrt{1-\left(\bhat/\ahat\right)^2}
\end{equation}
for prolates ($\ahat$\,$>$\,$\bhat$)
where $\ehat$ is the eccentricity,
and
\begin{equation}
L_{\|} = \frac{1+\ehat^2}{\ehat^2}\left(1-\frac{1}{\ehat}
      \arctan \ehat\right) ~~,~~
\ehat = \sqrt{\left(\bhat/\ahat\right)^2-1}
\end{equation}
for oblates ($\ahat$\,$<$\,$\bhat$).
For spheres $L_{\|}$\,=\,$L_{\bot}$\,=\,1/3 and $\ehat$\,=\,0. 
For extremely elongated prolates or ``needles''
($\ahat$\,$\gg$\,$\bhat$), it is apparent 
$C_{\rm abs}^{\bot}$\,$\ll$\,$C_{\rm abs}^{\|}$, 
we thus obtain
\begin{equation}\label{eq:needle}
\cabs/V \approx \frac{\omega}{3c}\frac{\epsilonim}
{\left[L_{\|}(\epsilonre-1) + 1\right]^2 
+ \left(L_{\|}\epsilonim\right)^2} 
\end{equation} 
where $L_{\|}$\,$\approx$\,$\left(\bhat/\ahat\right)^2\ln(\ahat/\bhat)$.
For dielectric needles, $\cabs \propto \omega\epsilonim
\propto \lambda^{-2}$ at long wavelengths 
since $L_{\|}(\epsilonre-1) + 1 \gg L_{\|}\epsilonim$
(see \cite{Li_needles_2003});
for {\it metallic} needles, for a given value of $\epsilonim$ 
one can always find a sufficiently long needle with 
$L_{\|}\epsilonim \simlt 1$ and
$L_{\|}(\epsilonre-1) \ll 1$ so that 
$\cabs \propto \omega\epsilonim \propto \sigma$
which can be very large
(see \cite{Li_needles_2003}).
Because of their unique optical properties,
metallic needles with
high electrical conductivities (e.g. iron needles,
graphite whiskers) are resorted to 
explain a wide variety of astrophysical phenomena:
(1) as a source of starlight opacity to create 
    a non-cosmological microwave background by 
    the thermalization of starlight in 
    a steady-state cosmology \cite{Hoyle_Wick_1988};
(2) as a source of the grey opacity needed to explain 
    the observed redshift-magnitude relation of Type Ia
    supernovae without invoking a positive cosmological 
    constant \cite{Aguirre_1999};
(3) as the source for the submm excess observed 
    in the Cas~A supernova remnant 
    \cite{Dwek_SN_2004}; and
(4) as an explanation for the flat 3--8$\mum$ extinction 
    observed for lines of sight toward the Galactic Center 
    and in the Galactic plane \cite{Dwek_extinction_2004}.
However, caution should be taken in using Eq.\ref{eq:needle}
(i.e. the Rayleigh approximation)
since the Rayleigh criterion $2\pi \ahat |m|/\lambda$\,$\ll$\,1 
is often not satisfied for highly conducting needles
(see \cite{Li_needles_2003}).\footnote{%
  The ``antenna theory'' has been applied 
  for conducting needle-like dust to estimate 
  its absorption cross sections \cite{Wright_1982}.
  Let it be represented by a circular cylinder
  of radius $\ahat$ and length $l$ ($\ahat$\,$\ll$\,$l$).
  Let $\rhoR$ be its resistivity.
  The absorption cross section is given by 
  $C_{\rm abs} = (4\pi/3c)(\pi \ahat^2 l/\rhoR)$,
  with a long wavelength cutoff of
  $\lambda_{\rm o} = \rhoR c (l/\ahat)^2/\ln(l/\ahat)^2$,
  and a short-wavelength cutoff of
  $\lambda_{\rm min} \approx  
  (2\pi c m_{e})/(\rhoR n_e e^2)$,
  where $m_e$, $e$, and $n_e$ are respectively
  the mass, charge, and number density of 
  the charge-carrying electrons. 
  }

In astronomical spectroscopy modeling, the continuous 
distribution of ellipsoid (CDE) shapes has been widely 
used to approximate the spectra of irregular dust grains
by averaging over all ellipsoidal shape parameters 
\cite{BH_1983}. In the Rayleigh limit, this approach, 
assuming that all ellipsoidal shapes are equally probable, 
has a simple expression for the average cross section 
\beq
\langle \cabs/V \rangle = \frac{\omega}{c} 
{\rm Im}\left\{\frac{2\varepsilon}{\varepsilon-1}\,
{\rm Log}\,\varepsilon\right\}~,~
{\rm Log}\,\varepsilon\,\equiv\,
\ln\sqrt{\epsilonre^2+\epsilonim^2}
+ i \arctan\left(\epsilonim/\epsilonre\right)
\eeq 
where ${\rm Log}\,\varepsilon$
is the principal value of the logarithm of $\varepsilon$.
The CDE approach, resulting in 
a significantly-broadened spectral band
(but with its maximum reduced),
seems to fit the experimental
absorption spectra of solids better than Mie theory.
Although the CDE may indeed represent a distribution 
of shape factors caused either by highly irregular 
dust shapes or by clustering of spherical grains into
irregular agglomerates, one should caution that 
the shape distribution of cosmic dust 
does not seem likely to resemble the CDE, 
which assumes that extreme shapes 
like needles and disks are equally probable.
A more reasonable shape distribution function would 
be like $dP/dL_{\|}$\,=\,$12 L_{\|}\,(1-L_{\|})^2$
which peaks at spheres ($L^{\|}$\,=\,1/3). 
This function is symmetric about spheres with respect to 
eccentricity $e$ and drops to zero for 
the extreme cases:
infinitely thin needles 
($e$\,$\rightarrow$\,1, $L_{\|}$\,$\rightarrow$\,0)
or infinitely flattened pancakes
($e$\,$\rightarrow$\,$\infty$, $L_{\|}$\,$\rightarrow$\,1).
Averaging over the shape distribution, 
the resultant absorption cross section is
$\cabs$\,=\,$\int_{0}^{1} dL_{\|} dP/dL_{\|} 
\cabs(L_{\|})$ where $\cabs(L_{\|})$ is 
the absorption cross section of a particular shape $L_{\|}$  
\cite{Ossenkopf_Henning_Mathis_1992, Li_Greenberg_Zhao_2002,
Li_Draine_2002_silicon}. 
Alternatively, Fabian et al. \cite{Fabian_etal_2001}
proposed a quadratic weighting for the shape distribution,
``with near-spherical shapes being most probable''.

\vspace{2mm}
When a dust grain is very large compared with the wavelength,
the electromagnetic radiation may be treated by geometric
optics: $\Qext$\,$\equiv$\,$\Cext/\Cgeo$\,$\rightarrow$\,2
if $x$\,$\equiv$\,$2\pi a/\lambda$\,$\gg$\,1 
{\it and} $|m-1|\,x$\,$\gg$\,1.\footnote{%
  At a first glance, $\Qext$\,$\rightarrow$\,2 appears to 
  contradict ``common sense'' by implying that a large grain
  removes twice the energy that is incident on it!
  This actually can be readily understood in terms
  of basic optics principles:
  (1) on one hand, all rays impinging on the dust 
  are either scattered or absorbed.  
  This gives rise to a contribution of $\Cgeo$
  to the extinction cross section.
  (2) On the other hand, all the rays in the field 
  which do not hit the dust give rise to 
  a diffraction pattern that is, by Babinet's principle,  
  identical to the diffraction through a hole of area $\Cgeo$. 
  If the detection excludes this diffracted light 
  then an additional contribution of $\Cgeo$ is made 
  to the total extinction cross section
  \cite{BH_1983}. 
  }
For these grains ($g$\,$\approx$\,1; $\cabs$\,$\approx$\,$\Cgeo$),  
the ratio of the radiation pressure to gravity 
$\beta_{\rm pr}$\,$\propto$\,$\cabs/a^3$\,$\propto$\,$1/a$.
This is demonstrated in Figure \ref{fig:beta_pr}. 
For dust with $x$\,$\gg$\,1 
{\it and} $|m-1|\,x$\,$\ll$\,1,
one can use the ``anomalous diffraction'' theory \cite{vandehulst_1957}. 
For dust with $|m-1|\,x$\,$\ll$\,1 {\it and} $|m-1|$\,$\ll$\,1, 
one can use the Rayleigh-Gans approximation\footnote{%
  The conditions for the Rayleigh-Gans approximation
  to be valid are $|m-1|$\,$\ll$\,1
  {\it and} $|m-1|\,x$\,$\ll$\,1. 
  The former ensures that the reflection from the surface 
  of the dust is negligible (i.e. the impinging light
  enters the dust instead of being reflected);
  the latter ensures that the phase of the incident wave 
  is not shifted inside the dust. 
  For sufficiently small scattering angles, 
  it is therefore possible for the waves scattered 
  throughout the dust to add coherently. 
  The intensity ($I$) of the scattered waves is 
  proportional to the number ($N$) of scattering sites squared:
  $I$\,$\propto$\,$N^2$\,$\propto$\,$\rho^2 a^6$
  (where $\rho$ is the mass density of the dust).
  This is why the X-ray halos 
  (usually within $\simali$1$^\circ$ surrounding 
   a distant X-ray point source; \cite{Overbeck_1965})
  created by the small-angle scattering of X-rays 
  by interstellar dust 
  are often used to probe the size 
  (particularly the large size end; 
  \cite{Draine_Tan_2003,Smith_etal_2002,Witt_etal_2001}), 
  morphology (compact or porous; 
  \cite{Mathis_etal_1995,Smith_Dwek_1998}), 
  composition, and spatial distribution of dust. 
  }
to obtain the absorption and scattering cross 
sections \cite{BH_1983,Laor_Draine_1993,vandehulst_1957}:
\beq
\Qabs \approx \frac{8}{3} {\rm Im}\left\{mx\right\}~~,~~
\Qsca \approx \frac{32\,|m-1|^2 x^4}{27+16 x^2}~~.
\eeq
It is important to note that the Rayleigh-Gans approximation 
is invalid for modeling the X-ray scattering by interstellar 
dust at energies below 1$\keV$.
This approximation systematically and substantially 
overestimates the intensity of the X-ray halo below $1\keV$
\cite{Smith_Dwek_1998}.

\vspace{-2mm}
\section{Scattering of Light by Dust: 
         Numerical Techniques
         \label{sec:Scat_Tech}}
\vspace{-2mm}
While simple analytic expressions exist for the scattering
and absorption properties of dust grains which are either
very small or very large compared to the wavelength of
the incident radiation (see \S\ref{sec:Scat_Approx}),
however, in many astrophysical applications we are
concerned with grains which are neither very small
nor very large compared to the wavelength.
Moreover, cosmic dust would in general be expected to 
have non-spherical, irregular shapes.
 
Our ability to compute scattering and absorption 
cross sections for nonspherical particles is extremely limited. 
So far, exact solutions of scattering problems exist only
for bare or layered spherical grains 
(``Mie theory''; \cite{BH_1983}),
infinite cylinders \cite{Lind_Greenberg_1966},
and spheroids 
\cite{Asano_Sato_1980,Asano_Yamamoto_1975,Vosh_Fara_1993}. 
The ``T-matrix'' (transition matrix) method, originally
developed by Barber \& Yeh \cite{Barber_Yeh_1975} 
and substantially extended by 
Mishchenko, Travis, \& Mackowski
\cite{Mishchenko_Travis_Mackowski_1996},
is able to treat axisymmetric 
(spheroidal or finite cylindrical) grains 
with sizes comparable to the wavelength.
The discrete dipole approximation (DDA), 
originally developed by Purcell \& Pennypacker
\cite{Purcell_Pennypacker_1973}
and greatly improved by Draine \cite{Draine_DDA_1988}, 
is a powerful technique for irregular heterogeneous grains 
with sizes as large as several times the wavelength. 
The VIEF (volume integration of electric fields) method 
developed by Hage \& Greenberg \cite{Hage_Greenberg_1990}, 
based on an integral representation of Maxwell's equations, 
is physically similar to the DDA method.
The microwave analog methods originally developed by
Greenberg, Pedersen \& Pedersen 
\cite{Greenberg_Pedersen_Pedersen_1960}
provide an effective experimental approach to complex particles
\cite{Gustafson_Kolokolova_1999}.

Although interstellar grains are obviously 
non-spherical as evidenced by the observed polarization 
of starlight, the assumption of spherical shapes 
(together with the Bruggeman or the Maxwell-Garnett 
effective medium theories for inhomogeneous grains; 
\cite{BH_1983}) is usually sufficient in modeling 
the interstellar absorption, scattering 
and IR (continuum) emission. 
For IR polarization modeling, 
the dipole approximation for spheroidal grains 
is proven to be successful in many cases.
  
The DDA method is highly recommended for studies 
of inhomogeneous grains and irregular grains 
such as cometary, interplanetary, 
and protoplanetary dust particles
which are expected to have a porous aggregate structure.
Extensive investigations using the DDA method have been 
performed for the scattering, absorption, 
thermal IR emission, and radiation pressure 
properties of fluffy aggregated dust
(e.g. see \cite{Andersen_etal_2006,
Gustafson_Kolokolova_1999,
Kimura_et al_2002,
Kimura_et al_2003,Kimura_et al_2006,
Kohler_etal_2006,Kohler_etal_2007,
Kolokolova_etal_2004,
Kozasa_etal_1992,Kozasa_etal_1993,
Lasue_Levasseur_2006,Levasseur_etal_2007,
Mann_etal_1994,Mann_etal_2004,
Mukai_etal_1992,
Okada_etal_2006,
Wilck_Mann_1996,
Wolff_etal_1998,
Xing_Hanner_1999,
Yanamandra_Hanner_1999}).

\vspace{3mm}
{\noindent {\bf Acknowledgements}~~ I thank Drs. I. Mann
and T. Mukai for their very helpful comments and support.
Partial support by a Chandra Theory program and 
HST Theory Programs is gratefully acknowledged.}



\printindex
\end{document}